\documentstyle[12pt,aps,epsfig]{revtex}
\newcommand{\be}{\begin{equation}}
\newcommand{\ee}{\end{equation}}
\newcommand{\bea}{\begin{eqnarray}}
\newcommand{\eea}{\end{eqnarray}}

\def\Journal#1#2#3#4{{#1} {\bf #2}, #3 (#4)}

\def\NP{{Nucl. Phys.}}
\def\PL{{Phys. Lett.} B}
\def\PR{{Phys. Rep.}}
\def\PRL{Phys. Rev. Lett.}
\def\PRD{{Phys. Rev.} D}
\def\PRC{{Phys. Rev.} C}
\def\ZPA{{Z. Phys.} A}
\def\ZPC{{Z. Phys.} C}
\def\JPG{{J. Phys.} G}
\def\HIP{{Heavy Ion Phys.}}
\def\MPLA{{Mod. Phys. Lett. A}}

\def\EPJ{{Eur. Phys. J.}}
\def\CPC{{Comput. Phys. Comm.}}

\begin{document}
\title{Collective dynamics in highly relativistic heavy-ion
collisions\footnote[1]{Supported by the German Academic Exchange
Service (DAAD).}}
\author{A. Dumitru}
\address{Physics Department, Yale University\\
P.O.\ Box 208124, New Haven, CT 06520, USA\\}
\author{D.H. Rischke}
\address{RIKEN-BNL Research Center, Brookhaven National Laboratory\\
Upton, NY 11973, USA}
\date{\today}
\maketitle   
\begin{abstract}
Hydrodynamics with cylindrical symmetry in transverse direction
and longitudinal scaling flow is employed to calculate the
transverse momentum spectra of various hadrons and clusters 
(e.g.\ $\pi$, $K$, $N$, $\Phi$, $\Lambda$, $d$, $He$) in central heavy-ion
collisions at CERN-SPS and BNL-RHIC energies up to $p_T=4$~GeV.
We discuss the sensitivity of these spectra with respect to
the initial transverse density profile as well as to 
the choice of ``freeze-out'' hypersurface.
For $\sqrt{s}=18A$~GeV (SPS energy) and $p_T<2$~GeV, overall good
agreement of the $p_T$ distributions with data is found when
freeze-out occurs along the $T=130$~MeV isotherm.
Even high-$p_T$ neutral pion data can 
be described for a particular choice of the initial transverse
density profile. It is shown that the average transverse 
velocity $\langle v_T\rangle$ of heavy hadrons and hadronic clusters 
is a good measure for the collective flow velocity. The latter is found 
to be rather similar for SPS and RHIC
energies, due to the ``stall'' of the flow within the long-lived mixed
phase at RHIC. In case of thermalization and
hydrodynamical expansion, the mean transverse momentum
$\langle p_T\rangle$ increases linearly with the
hadron mass. In contrast, the string model {\small FRITIOF~7.02}, which does
not account for rescattering of secondary hadrons, predicts a strong
dependence of $\langle p_T\rangle$ on the quark composition of the hadron.
Due to the different sensitivity to hard processes, hadrons with charm
(anti-)quarks acquire significantly more transverse momentum than hadrons
without $c$ (or even without $s$) quarks.
\end{abstract}

\section{Introduction}
A variety of experimental data for rapidity and transverse momentum
spectra of various hadrons and clusters 
\cite{NA49h-,NA49phi,AGSflow,BMS,vTExp} in
relativistic heavy-ion collisions
at the BNL-AGS and CERN-SPS indicate significant
collective (longitudinal and transverse) expansion of the produced hot and
dense fireball~\cite{Larisa,RaffiPRL,Marburg,JSol,SHR,BSch,ACRS,Hecke}.
Also at the much higher center-of-mass energy
that will be reached at the BNL-RHIC, it is commonly expected
that the parton density at midrapidity in central collisions is high enough
to lead to rapid thermalization~\cite{T0,PCM,MiniJet}.

In the present paper we discuss the collective evolution of the
system following thermalization within
a hydrodynamical model. We compute the transverse
momentum spectra at midrapidity $dN/d^2p_Tdy|_{y=0}$,
the mean transverse momenta $\langle p_T\rangle$, and the mean
transverse velocities $\langle v_T\rangle$ of various
hadron species for $Pb+Pb$ at CERN-SPS energy 
($\sqrt{s}=18A$~GeV, entropy per net baryon $s/\rho_B \approx 40$)
and $Au+Au$ at BNL-RHIC energy ($\sqrt{s}=200A$~GeV, $s/\rho_B
\approx 200$). Finally, we also calculate the $\langle p_T\rangle$
of various hadrons in $Au+Au$ at RHIC energy within a string model
without rescattering of produced particles ({\small FRITIOF~7.02}~\cite{FRI})
and compare its predictions to those of hydrodynamics.

\section{Model Description} \label{ModDescr}
\subsection{Scaling Hydrodynamics}
Hydrodynamics is defined by (local) energy-momentum and net charge
conservation~\cite{Landau},
\be \label{Hydro}
\partial_\mu T^{\mu\nu}=0 \quad,\quad
\partial_\mu N_i^{\mu}=0\quad.
\ee
$T^{\mu\nu}$ denotes the energy-momentum tensor, and $N_i^{\mu}$ the net
four-current of the $i$th conserved charge. We will explicitly consider only
one such conserved charge, the net baryon number. We implicitly assume
that all other charges which are conserved on strong-interaction time scales,
e.g.\ strangeness, charm, and electric charge, vanish locally. The
corresponding four-currents are therefore identically zero, cf.\
eq.~(\ref{idfluid}), and the conservation equations are trivial.

For ideal fluids, the energy-momentum tensor and the net baryon current
assume the simple form~\cite{Landau}
\be \label{idfluid}
T^{\mu\nu}=\left(\epsilon+p\right) u^\mu u^\nu -p g^{\mu\nu}
\quad,\quad
N_B^{\mu}=\rho_B u^\mu \quad,
\ee
where $\epsilon$, $p$, $\rho_B$ are energy density, pressure, and net baryon
density in the local rest frame of the fluid, which is defined by
$N_B^\mu=(\rho_B,\vec{0})$. $g^{\mu\nu}={\rm diag}(+,-,-,-)$ is the metric
tensor, and $u^\mu=\gamma(1,\vec{v})$ the four-velocity of the fluid
($\vec{v}$ is the three-velocity and $\gamma=(1-\vec{v}^2)^{-1/2}$ the
Lorentz factor). The system of partial differential
equations~(\ref{Hydro}) is closed by choosing an equation of state (EoS)
in the form $p=p(\epsilon,\rho_B)$, cf.\ below.

For simplicity, we assume cylindrically symmetric transverse expansion with a
longitudinal scaling flow profile, $v_z=z/t$~\cite{Bj}.
At $z=0$, equations~(\ref{Hydro}) reduce to
\bea \label{2Deq}
\partial_t E+\partial_T \left[\left(E+p\right)v_T\right] &=&
  - \left(\frac{v_T}{r_T}+\frac{1}{t}\right)\left(E+p\right)\quad,\nonumber\\
\partial_t M+\partial_T \left(Mv_T+p\right) &=&
  - \left(\frac{v_T}{r_T}+\frac{1}{t}\right) M\quad,\\
\partial_t R+\partial_T \left(Rv_T\right) &=&
  - \left(\frac{v_T}{r_T}+\frac{1}{t}\right) R\quad, \nonumber
\eea
where we defined $E\equiv T^{00}$, $M\equiv T^{0T}$, and $R\equiv N_B^0$.
In the above expressions, the index $T$ refers to the transverse component
of the corresponding quantity.

The set of equations~(\ref{2Deq}) describes the evolution in the $z=0$
plane. Due to the assumption of longitudinal scaling,
the solution at any other $z\neq0$ can be simply obtained by a
Lorentz boost. The above equations also imply
\be \label{peta}
\left.\frac{\partial p}{\partial\eta}\right|_{\tau,r_T}=0 \quad,
\ee
where $\eta\equiv{\rm Artanh}~v_z$ and
$\tau\equiv\sqrt{t^2-z^2}$. This means that on proper-time hyperbolas
pressure gradients in rapidity direction vanish, and there is no flow
between adjacent infinitesimal rapidity slices. However, only for net
baryon free matter, $\rho_B\equiv0$, does this automatically also mean
that the temperature $T$ is independent of the longitudinal fluid rapidity
$\eta$. In the case $\rho_B\neq0$ equation~(\ref{peta}) only demands
\be \label{Tmueta}
\left. s\frac{\partial T}{\partial\eta}\right|_\tau +
\left. \rho_B\frac{\partial \mu_B}{\partial\eta}\right|_\tau=0\quad.
\ee
$s$ and $\mu_B$ denote entropy density and baryo-chemical potential,
respectively. If other charges like strangeness or electric charge are locally
nonvanishing, additional terms appear.
Equation (\ref{Tmueta}) does not imply that the rapidity
distribution of produced particles is flat (i.e.\ independent of rapidity)
or that the rapidity distributions of various species of hadrons, e.g.\
pions, kaons, and nucleons, are similar. {\em Any rapidity-dependent}
$T$ and $\mu_B$ that satisfy eq.~(\ref{Tmueta}) are in agreement with
energy-momentum and net baryon number conservation, as well as with
longitudinal scaling flow $v_z=z/t$. In this paper, however, we do not
explore the rapidity dependence of the particle spectra, and thus, as
discussed below, are not required to specify
$T$ and $\mu_B$ as functions of $\eta$.

The hydrodynamical equations of motion are solved on a discretized
space-time grid ($\Delta r_T=R_T/100=0.06$~fm, $\Delta \tau=0.99\Delta r_T$)
employing the RHLLE algorithm as described and tested in~\cite{RiGy,RiBe}.
We have checked that the algorithm conserves total
energy and baryon number to within $1\%$ and that profiles of rarefaction
and shock waves are reproduced accurately for various initial
conditions~\cite{RiBe,RiGy2}.

\subsection{Equation of State}
To close the system of coupled equations of hydrodynamics, an equation of
state (EoS) has to be specified. Lattice QCD predicts a phase transition
from ordinary nuclear matter to a so-called quark-gluon plasma (QGP) at
a critical temperature of $T_C=140-160$~MeV~\cite{Laerm} (for $\rho_B=0$).
We model the high-temperature phase as an ideal gas of $u$, $d$,
$s$ quarks (with masses $m_u=m_d=0$, $m_s=150$~MeV), and gluons, employing
the well-known
MIT bag model EoS~\cite{MIT}. In the low-temperature region we assume an
ideal hadron gas that includes the well-established (strange and nonstrange)
hadrons up to masses of $2$~GeV (as listed in~\cite{Winck}).
The phase coexistence region 
is constructed employing Gibbs' conditions of phase equilibrium. The bag
parameter of $B=380$~MeV/fm$^3$ is chosen to yield the critical temperature
$T_C\approx160$~MeV.

\subsection{Initial Conditions}
Equation (\ref{peta}) suggests that it is convenient to specify 
the initial conditions in
scaling hydrodynamics on a proper-time hyperbola $\tau=\tau_i$.
Apart from $\tau_i$, we also have to specify the entropy per net baryon,
$s/\rho_B$, and the net baryon rapidity density at midrapidity, $dN_B/dy$.
The initial energy and net baryon densities as well
as the temperature and the chemical potentials are then determined
by the EoS. After specifying an initial transverse density distribution, 
the solution of relativistic ideal hydrodynamics 
is uniquely determined in the forward lightcone.

For collisions at SPS energy we assume that hydrodynamic flow starts at
proper time $\tau_i=1$~fm/c. This is a value conventionally assumed in the
literature, cf.\ e.g.~\cite{Bj}. Note that the production time of the
partons in the central region might be smaller,
$\simeq 0.2-0.3$~fm, cf.\ e.g.~\cite{KMcLS}. However, we assume that
one to two further collisions per particle are necessary for thermalization
and onset of collective behavior.

We further assume $dN_B/dy=80$, which
(after resonance decays) results in $\approx60$ net nucleons at midrapidity.
If we simply scale by 1/2 (at high energy the initial isospin asymmetry is
transferred to the pions and the nucleons become almost isospin symmetric),
the net proton multiplicity at midrapidity agrees with preliminary
NA49 data~\cite{NA49h-}.

The specific entropy in central heavy-ion collisions at SPS energies is
$s/\rho_B=40-50$, cf.\ e.g.\ refs.~\cite{BMS,3fluid,CSCG} and
references therein. With this entropy per net baryon most hadron
multiplicities
as measured at CERN-SPS can be well described, e.g.\ the multiplicity of
pions in central $Pb+Pb$ at midrapidity, $dN_\pi/dy\approx600$~\cite{NA49h-}.

The initial energy and net baryon densities follow as
$\epsilon_i=5.3$~GeV/fm$^3$ and $\rho_B=4.5\rho_0$, respectively.
The initial temperature and quark-chemical potentials are
$T_i=216$~MeV, and $\mu_q=167$~MeV, $\mu_s=0$, respectively.
An estimate of the initial energy density could also be obtained
using the Bjorken formula
\be \label{BjTE}
\epsilon_i = \frac{dE_T/dy}{\tau_i \pi R_T^2}\approx3~{\rm GeV/fm}^3\quad,
\ee
where $dE_T/dy=405$~GeV~\cite{PbPbTrEnergy} is the measured transverse
energy in the most central $Pb+Pb$ events. Since this simple formula does
not account for the work performed by expansion, cf.\
e.g.~\cite{SHR}, it predicts an initial energy density that is almost a factor
of two below that assumed in our calculation.
We obtain $dE_T/dy=412$~GeV on the hadronization hypersurface, and $dE_T/dy=
406$~GeV on the $T=130$~MeV isotherm.

Let us now turn to the initial conditions for central $Au+Au$ collisions
at RHIC energy. Due to the higher parton
density at midrapidity (as compared to collisions at SPS energy),
thermalization may be reached earlier at RHIC. According to various
studies~\cite{T0,PCM,MiniJet}, thermalization might occur within $\sim
0.5$ fm. Following~\cite{RiGy} we will assume $\tau_i=R_T/10=0.6$~fm 
for $Au$ nuclei.

Various microscopic models,
e.g.\ {\small PCM}~\cite{PCM}, {\small RQMD~1.07}~\cite{RQMD107},
{\small FRITIOF~7.02}~\cite{FRIT}, and {\small HIJING/B}~\cite{HIJING},
predict a net baryon rapidity density of $dN_B/dy\approx20-35$ and 
specific entropy of $s/\rho_B\approx150-250$ in central $Au+Au$ at
$\sqrt{s}=200A$~GeV at midrapidity. Perturbative QCD calculations of minijet
production in leading logarithm approximation with a transverse momentum
cutoff of $p_0=1$~GeV yield $s/\rho_B=150$, if a soft component with
$s/\rho_B=50$ is assumed~\cite{MiniJet}.

We will employ
$dN_B/dy=25$ and $s/\rho_B=200$. For our choice of $\tau_i$ and equation
of state, the resulting energy and baryon densities at
midrapidity are $\epsilon_i=17$~GeV/fm$^3$ and
$\rho_i=2.3\rho_0$. The initial temperature and quark-chemical potentials
follow as $T_i=300$~MeV, $\mu_q=47$~MeV, $\mu_s=0$, respectively.
The higher specific entropy and lower net baryon number in $Au+Au$ collisions
at RHIC energy as compared to SPS energy
enhance the pion multiplicity at midrapidity by
$\approx50\%$~\cite{DumSpi}. Also, antibaryon to baryon ratios at midrapidity
increase from $\approx 0.1$ to $\ge 0.5$~\cite{DumSpi}.

The total transverse energy at midrapidity on the hadronization hypersurface
is $dE_T/dy=641$~GeV, and
decreases to 627~GeV on the $T=130$~MeV isotherm. Thus, the isentropic
expansion reduces the energy at midrapidity by almost a factor of
two ($dE_T/dy(\tau_i)=1153$~GeV).

In an individual event, the transverse flow velocity field can be rather
complicated~\cite{HotSpot}, however, on average over many events it
vanishes at $\tau=\tau_i$. The radial energy
and net baryon densities are taken to be proportional to
$\Theta(R_T-r_T)$. $R_T=6$~fm denotes the nuclear radius. To show the effect
of initial pressure gradients
on collective particle production at high $p_T$ we will also employ
the ``wounded nucleon'' distribution
\bea 
\epsilon(\tau_i,r_T) &=& {\epsilon_i}f\left(r_T\right) \quad,\nonumber\\
\rho_B(\tau_i,r_T) &=& {\rho_i}f\left(r_T\right) \quad, \label{wn_distrib}
\eea
with
\be
f\left(r_T\right) = \frac{3}{2}\sqrt{1-\frac{r_T^2}{R_T^2}} \quad.
\ee
$\epsilon_i$, $\rho_i$ denote the energy and net baryon densities as
specified above for SPS and RHIC energies, respectively.

\subsection{Calculation of Particle Momentum Spectra}
It is generally unreasonable to assume that the fluid-dynamical description
is valid in the whole forward lightcone. In some space-time regions
the reaction rates between the particles are too low to maintain local
thermal and/or chemical equilibrium. In hydrodynamical calculations
one usually assumes that the transition between the fluid and the
free streaming regime occurs within a space-time volume
which is commonly approximated by a three-dimensional surface in four
dimensional space-time~\cite{Landau,FOgen}.
One then has to determine the parameters of the free-streaming
gas, characterizing its momentum-space distribution, from the properties
of the fluid (i.e.\ velocity, temperature, etc.) on the assumed freeze-out
hypersurface. From the theoretical point of view, this complicated
problem is not yet satisfactorily solved \cite{AltFO}.
Problems may arise, in particular, if
the momentum-space distribution functions on the two sides of the
freeze-out hypersurface (fluid and free-streaming gas) differ.
Also, on time-like parts
of the freeze-out hypersurface (i.e.\ with space-like normal) negative
contributions to the spectrum of frozen-out particles may arise. For high
flow velocities through the hypersurface, however, these contributions
are negligible.

Several attempts to understand and model the freeze-out can be found in
the literature~\cite{FOgen,AltFO,HungS}.
We will employ the description of Cooper and Frye~\cite{CF},
which has been used to calculate rapidity and transverse momentum
spectra at AGS and SPS energies by several
authors~\cite{Marburg,JSol,SHR,BSch,ACRS,HungS,CFRHI,CRS}.
For a cylindrically symmetric transverse expansion exhibiting longitudinal
scaling flow the particle momentum spectra are given
by
\be \label{CFSpectra}
\frac{dN}{d^2p_Tdy}= \int\limits_0^1 d\zeta \int\limits_0^{2\pi} d\phi
\int\limits_{-\infty}^\infty d\eta \,\, r_T\,\tau\left( p_T\cos\phi
\frac{d\tau}{d\zeta} -m_T\cosh(y-\eta)\frac{dr_T}{d\zeta}\right)\,
f\left(p_\mu u^\mu\right)\quad,
\ee
where $p^\mu=(m_T\cosh y,p_T\sin\chi,p_T\cos\chi,m_T\sinh y)$, and
$u^\mu=\gamma_T (\cosh\eta,v_T\sin(\phi-\chi),v_T\cos(\phi-\chi),
\sinh\eta)$ denote the particle four-momentum and the fluid four-velocity,
respectively. Thus, the direction of the particle momentum in the
transverse plane is determined by the angle $\chi$, while the relative
angle between $\vec{p}_T$ and the transverse flow velocity, $\vec{v}_T$,
is denoted by $\phi$. $\zeta\in [0,1]$ parametrizes the hypersurface 
(counter-clockwise) on which the momentum distribution of particles is to be
calculated, such that $r_T(\zeta)$ and $\tau(\zeta)$ specify the space-time
points on the hypersurface. Since we
focus on particles with $m_T\gg T_{fo}$, we approximate the phase-space
distribution function $f$ by a Boltzmann distribution, and also
\be\label{apprdelt}
\exp\left(-\frac{\gamma_T m_T}{T}\cosh\left(y-\eta\right)\right)
\approx \sqrt{\frac{2\pi T}{\gamma_T m_T}} 
\delta\left(y-\eta\right)
\exp\left(-\frac{\gamma_T m_T}{T}\right)\quad.
\ee
The thermal smearing of the rapidity distribution is thus approximated by
the square-root factor. The quality of this approximation is less good
for low-$p_T$ pions which are therefore not considered here.

For SPS energies, we calculate the transverse momentum spectra on the
$T=130$~MeV isotherm. This turns out to reproduce the measured spectra of
$\pi$, $p$, $\phi$, $\Lambda$, and $d$ reasonably well, cf.\ below. Also,
this choice is in line with freeze-out temperatures extracted by
others~\cite{NA49phi,BMS,vTExp,JSol,SHR,BSch}.

Even without a detailed calculation of the reaction rates between various
hadron species as a function of temperature and chemical potentials
(cf.~\cite{PPVW} for such a study and~\cite{HungS} for a first step to
include it into the fluid-dynamical model) one might 
speculate that the freeze-out temperature might be higher at RHIC 
than at SPS. The reason is that the lower net
baryon density has to be compensated for by an increase in the meson density,
in order that the reaction rates be large enough to maintain local
thermodynamical
equilibrium. We shall thus calculate the particle momentum
distributions at RHIC on the hypersurface where the fluid enters the
hadronic phase\footnote{Note that, for the EoS used here, hadronization
proceeds via a shock wave~\cite{RiBe,Bug}. In practice, we evaluate
eq.~(\ref{CFSpectra}) on the hypersurface corresponding to a small negative
$\lambda=-0.05$. This ensures that the fluid properties
(velocity, temperature etc.) are those of {\em hadronic\/} matter,
located either on or behind the shock front.}, $\lambda=0$. ($\lambda\equiv
V^{QGP}/V^{tot}$ denotes the fraction of quark-gluon plasma within
the mixed phase.)
The results are not changed if a $T=160$~MeV isotherm is employed instead.
To explore the range of uncertainty, however, we shall also show
results for freeze-out at $T=130$~MeV.

\section{Results}
\subsection{Space-Time Evolution}
\begin{figure}[htp]
\centerline{\hbox{\epsfig{figure=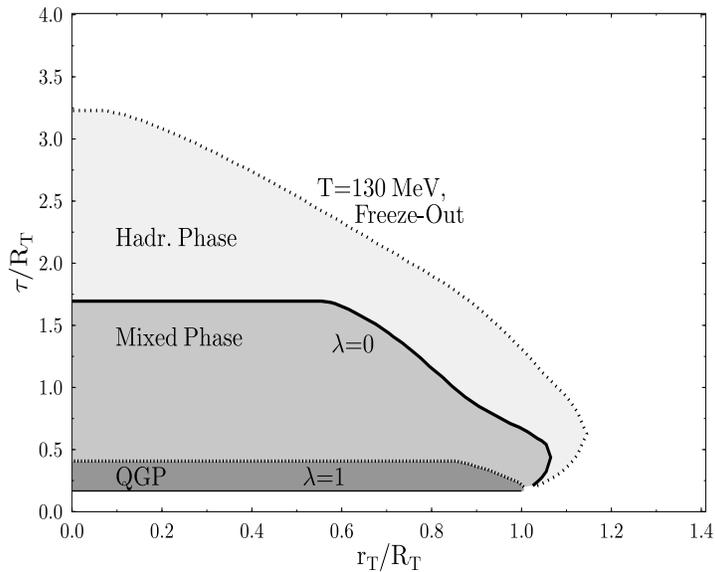,height=8cm,width=10cm}}}
\caption{Hypersurfaces corresponding to $\lambda=1$ (boundary
between pure QGP and mixed phase), $\lambda=0$ (boundary
between mixed phase and pure hadron phase), and the freeze-out isotherm
$T=130$~MeV; for central $Pb+Pb$ collisions at SPS.}
\label{hyperSPS}
\end{figure}
\begin{figure}[htp]
\centerline{\hbox{\epsfig{figure=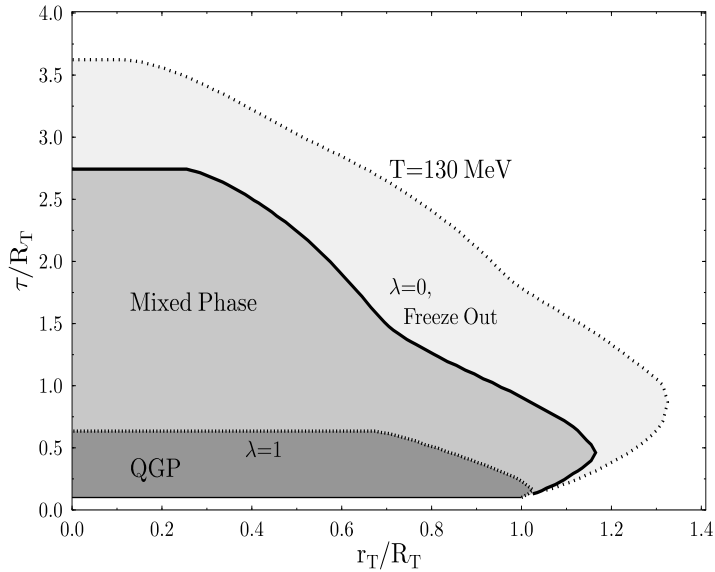,height=8cm,width=10cm}}}
\caption{Hypersurfaces corresponding to $\lambda=1$ (boundary
between QGP and mixed phase), and the freeze-out hypersurface
$\lambda=0$ (boundary between mixed phase and pure hadron phase);
for central $Au+Au$ collisions at RHIC.}
\label{hyperRHIC}
\end{figure}

  
Figures \ref{hyperSPS} and \ref{hyperRHIC} show the hypersurfaces where the
pure QGP and the mixed phase end, respectively, as well as the $T=130$~MeV
isotherm. One observes a qualitative change in the space-time evolution.
When comparing the evolution at SPS and RHIC, one finds that due to the high
initial entropy density the QGP phase lives almost twice as long at RHIC than
at SPS. Therefore, the transverse rarefaction wave can propagate further
into the quark-gluon fluid and accelerate it. This leads to larger transverse
flow of matter entering the mixed phase at RHIC than at SPS. Due to this
effect the hadronization surface ($\lambda=0$) extends to larger
$r_T$. Note also that at SPS the fluid spends $\approx50\%$ of its
lifetime in the hadronic phase, while at RHIC QGP and mixed phase occupy a
much larger space-time volume. For comparison, we have also shown the
$T=130$~MeV isotherm for $Au+Au$ reactions at RHIC.

For our EoS, which contains all
well-established hadrons up to mass $2$~GeV, hadronization is completed at
$\tau=1.7R_T=10$~fm (SPS) resp.\ $\tau=2.7R_T=16$~fm (RHIC). The values of
the lifetime of the mixed phase obtained here agree with those predicted in
ref.~\cite{RiGy} for the case where the ratio of degrees of freedom in the
QGP and hadronic phase is 3, cf.\ Fig.~12c in~\cite{RiGy}. If only $\pi$ or
$\pi$, $\eta$, $\rho$, $\omega$ mesons are considered in
the hadron gas, complete hadronization takes considerably longer at RHIC,
$\tau=6R_T=36$~fm \cite{RiGy} and $\tau=30$~fm \cite{Alam}, respectively
(cf.\ also refs.~\cite{JSol,SHR,BSch,CRS}).
The back-reaction of the freeze-out on the dynamics of the fluid could
further speed up hadronization by up to $30\%$~\cite{DumSpi}.

\subsection{Transverse Momentum Spectra of Hadrons and Clusters}
\begin{figure}[htp]
\centerline{\hbox{\epsfig{figure=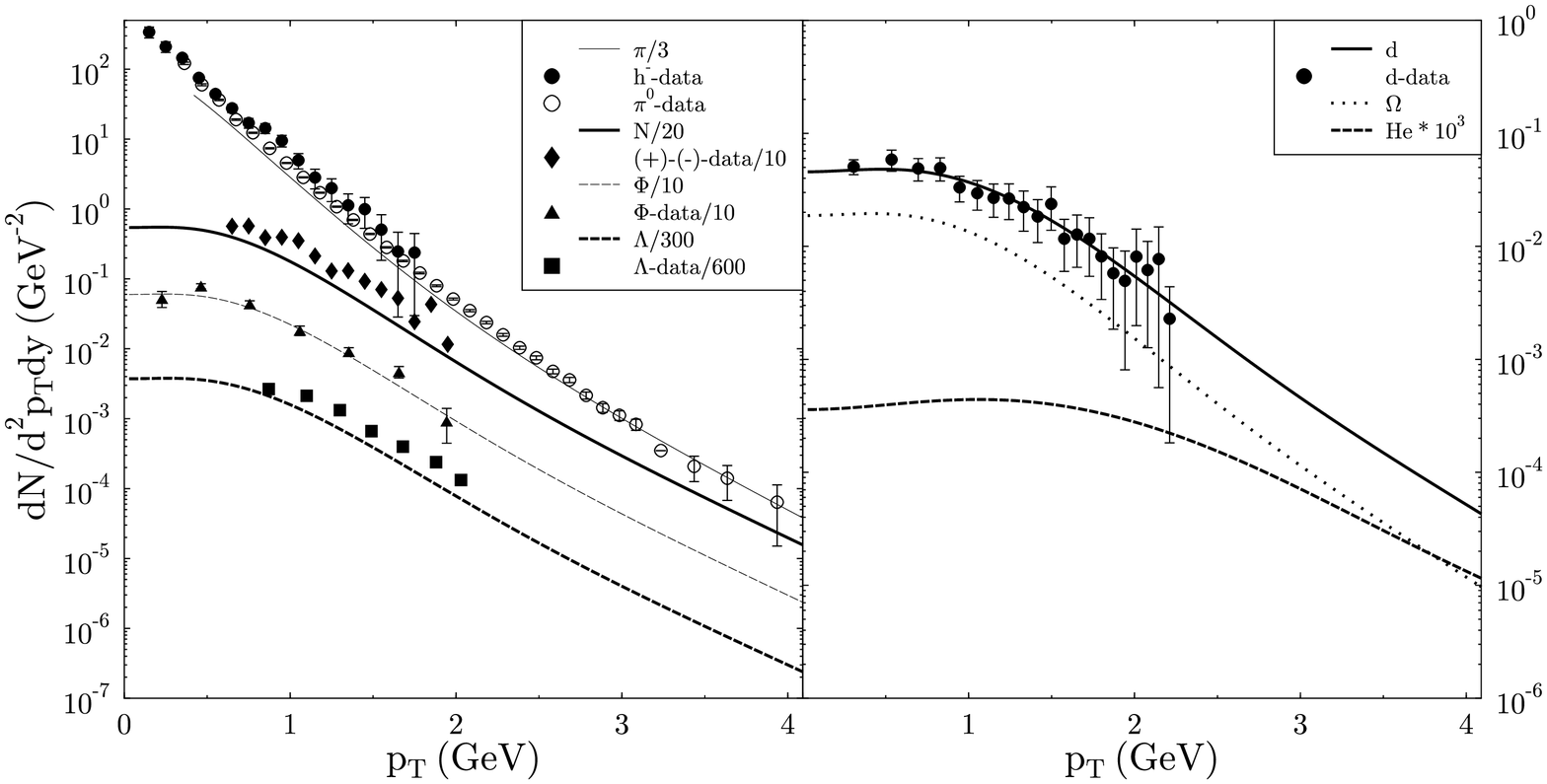,height=8cm,width=13cm}}}
\vspace*{1cm}
\caption{Transverse momentum spectra of direct thermal hadrons at midrapidity
(central $Pb+Pb$ collisions at SPS), calculated on the $T=130$~MeV isotherm.
A step function has been employed for the initial transverse energy and
baryon density profiles.}
\label{dndptSPS}
\vspace{.5cm}
\end{figure}  
Figure \ref{dndptSPS} compares our calculated $p_T$ spectra with $h^-$,
$(+)-(-)$~\cite{NA49h-}, $\phi$, $d$~\cite{NA49phi}, and
$\Lambda$~\cite{Bormann} data of NA49. The additional factor $1/2$ in the
$\Lambda$ data as compared to our calculation approximately
corrects for the $\Sigma^0$s, which can experimentally
not be separated from the $\Lambda$s. The preliminary $\pi^0$ data obtained
by the WA98 collaboration are also shown~\cite{WA98pi} (see also the data
obtained by the CERES collaboration~\cite{CERES}).
Both the calculated spectra as well as the experimental
data are absolute numbers without any arbitrary normalization factor.
The reason for the underpredicted multiplicities of pions, protons, and
lambdas is the fact that we have not included post-freeze-out resonance
decays\footnote{The total multiplicities, including feeding from resonance
decays and assuming chemical freeze-out at the phase boundary were calculated
e.g. in refs.~\cite{BMS,CSCG,DumSpi}.}. Also, we have not explored effects
of an isospin asymmetric initial state.

The {\em shapes} of the spectra, however, are well reproduced.
In particular, one can see the broadening of the $p_T$ distribution as
the particle mass increases, and the ``shoulder-arm'' structure that develops
in the spectra of the heavy clusters. This behavior emerges naturally in the
presence of collective transverse flow \cite{RaffiPRL,ACRS,StoOG,SPSvTth},
and has already been discovered experimentally at the BEVALAC~\cite{BEVflow}
and at the AGS~\cite{AGSflow}.

Note also that due to collective transverse flow the spectra at high $p_T$ are
not exponentials. The inverse slope of the thermal pions and nucleons in the
region $1$~GeV$<p_T<2$~GeV is $T^*_\pi=195$~MeV,
$T^*_N=220$~MeV, and increases to $T^*_\pi=255$~MeV,
$T^*_N=270$~MeV, respectively, at higher transverse
momentum, $2$~GeV$<p_T<3$~GeV. This effect is even stronger in $Au+Au$ at
RHIC, see below.  

The apparent temperature of high-$p_T$ pions agrees
well with the preliminary result of the WA98 collaboration,
$T^*_\pi=250$~MeV at $p_T=2$~GeV~\cite{WA98pi}. This holds also for the
multiplicity of $\pi^0$s with $p_T>2$~GeV at midrapidity; while we obtain
$0.32/3=0.11$ (excluding $\pi^0$s from post-freeze-out resonance decays)
the WA98 collaboration found $0.138\pm0.05$ (preliminary) $\pi^0$s in this
region of momentum space. In contrast to the arguments of
ref.~\cite{wangpi} it can thus not be excluded that the measured high-$p_T$
pions can be described within a fluid-dynamical model.

\begin{figure}[htp]
\centerline{\hbox{\epsfig{figure=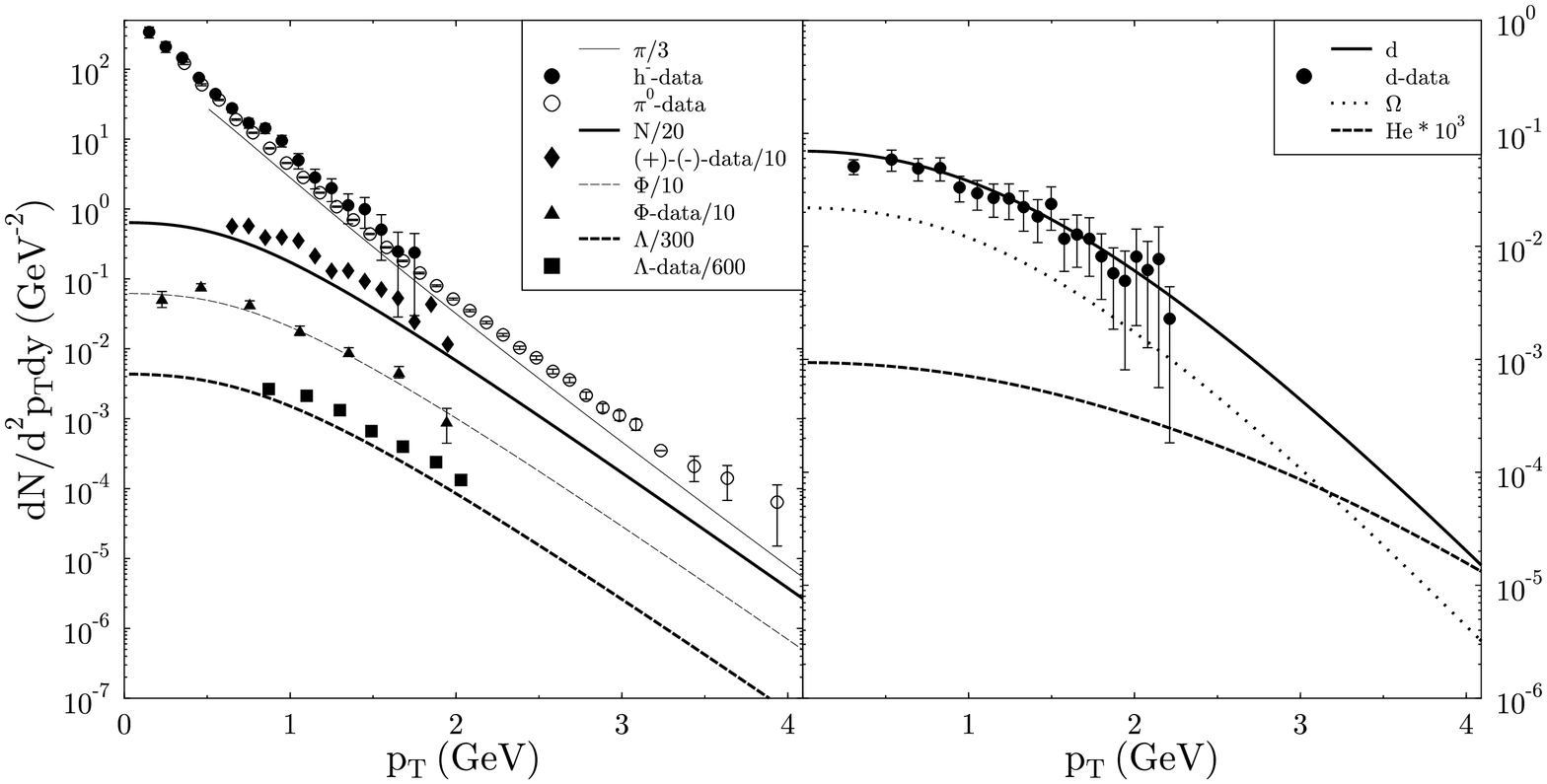,height=8cm,width=13cm}}}
\vspace*{1cm}
\caption{Transverse momentum spectra of direct thermal hadrons at midrapidity
(central $Pb+Pb$ collisions at SPS), calculated on the $T=130$~MeV isotherm.
A ``wounded nucleon'' distribution
has been employed for the initial transverse energy and baryon density.}
\label{dndptSPSwn}
\vspace{.5cm}
\end{figure}  
We emphasize, however, that collective
particle production at high $p_T$ shows some sensitivity to the initial
conditions.
Employing a ``wounded nucleon'' distribution, cf.\ eq.~(\ref{wn_distrib}),
the pion spectrum (on the $T=130$~MeV hypersurface)
above $p_T=1$~GeV is again closer to an exponential, 
cf.\ Fig.~\ref{dndptSPSwn}.
On the other hand, boost non-invariant initial
conditions~\cite{JSol,SHR,BSch}, or initial transverse velocity
gradients~\cite{HotSpot}, can again enhance collective transverse expansion.
In view of these uncertainties, we argue that collective hadron production
above $p_T=1$~GeV cannot be definitely excluded.

\begin{figure}[htp]
\centerline{\hbox{\epsfig{figure=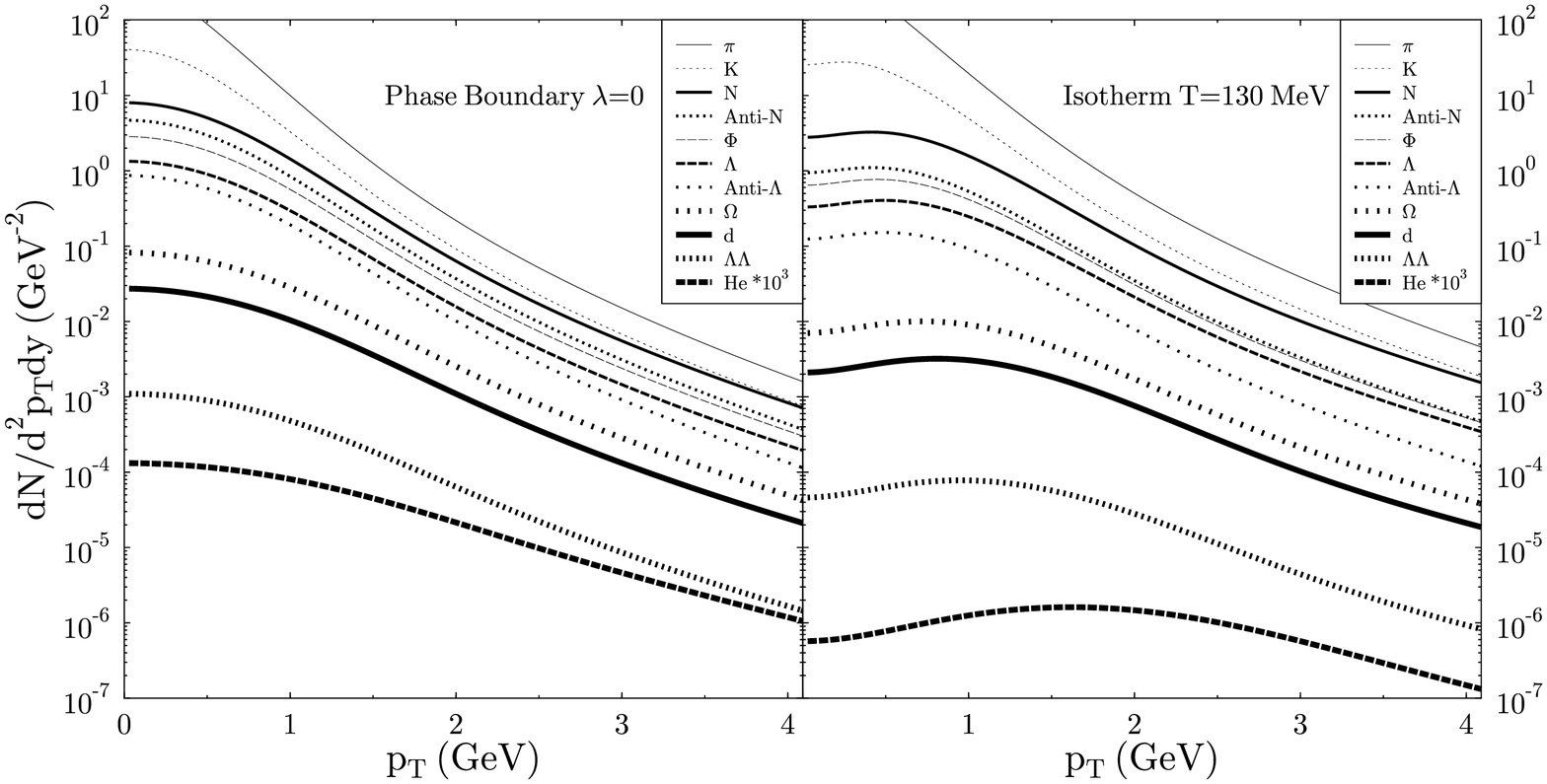,height=8cm,width=13cm}}}
\vspace*{1cm}
\caption{Transverse momentum spectra of direct thermal hadrons at midrapidity
(central $Au+Au$ collisions at RHIC), as calculated on the $\lambda=0$ and
$T=130$~MeV hypersurfaces, respectively. 
A step function has been employed for the initial transverse energy and
baryon density profiles.}
\label{dndptRHIC}
\vspace{.5cm}
\end{figure}
Figure \ref{dndptRHIC} shows the $p_T$ spectra for RHIC energy as calculated
on the $\lambda=0$ and $T=130$~MeV hypersurfaces.
The $\Lambda\Lambda$ is a hypothetical bound
state of two $\Lambda$ baryons for which we assume a binding energy of
$30$~MeV. If freeze-out occurred later, 
e.g.\ on the $T=130$~MeV isotherm like in $Pb+Pb$ at SPS, the
resulting spectra are of course ``stiffer'', since
collective flow is significantly stronger than on the
$\lambda=0$ hypersurface. This leads to much broader $p_T$ distributions for
the heavy hadrons. Observe also that the number of pions at $p_T=4$~GeV
increases by a factor of 3.

We would like to emphasize also another qualitative difference in the
dynamics at SPS and RHIC, where the space-time volume of the hadronic phase
might be smaller. At RHIC, the outer part of
the fluid ($r_T>0.7R_T$, cf.\ Fig.~\ref{hyperRHIC}) is strongly accelerated
by the rarefaction wave which moves into the QGP.
However, once longitudinal expansion has cooled the core of the fluid
into the phase coexistence region, the transverse 
rarefaction wave is stopped abruptly, since the velocity of sound in the 
mixed phase is small.
Thus, if the space-time volume of the purely hadronic phase is
small, the interior part of the fluid will remain more or less
at rest\footnote{Parametrization of collective transverse flow in terms of
a simple radial velocity profile~\cite{SPSvTth} is therefore not possible
on the $\lambda=0$ hypersurface. This is in contrast to freeze-out occurring
later, e.g.\ on the $T=130$~MeV isotherm.}.
We consequently expect two different slopes in the $p_T$ distributions.
This is confirmed by Fig.~\ref{dndptRHIC}, left panel. In the range
$1$~GeV$<p_T<2$~GeV, the apparent temperature of thermal pions and nucleons
is $T^*_\pi=220$~MeV and $T^*_N=233$~MeV, respectively; at higher $p_T$,
$2$~GeV$<p_T<3$~GeV, the inverse slopes increase by roughly $100$~MeV, to
$T^*_\pi=325$~MeV, and $T^*_N=335$~MeV~! Note that the difference
is much larger than in $Pb+Pb$ at SPS, although the
average transverse flow velocity (on the $\lambda=0$ hypersurface) is smaller
than on the $T=130$~MeV isotherm at SPS (see below).

\begin{figure}[htp]
\centerline{\hbox{\epsfig{figure=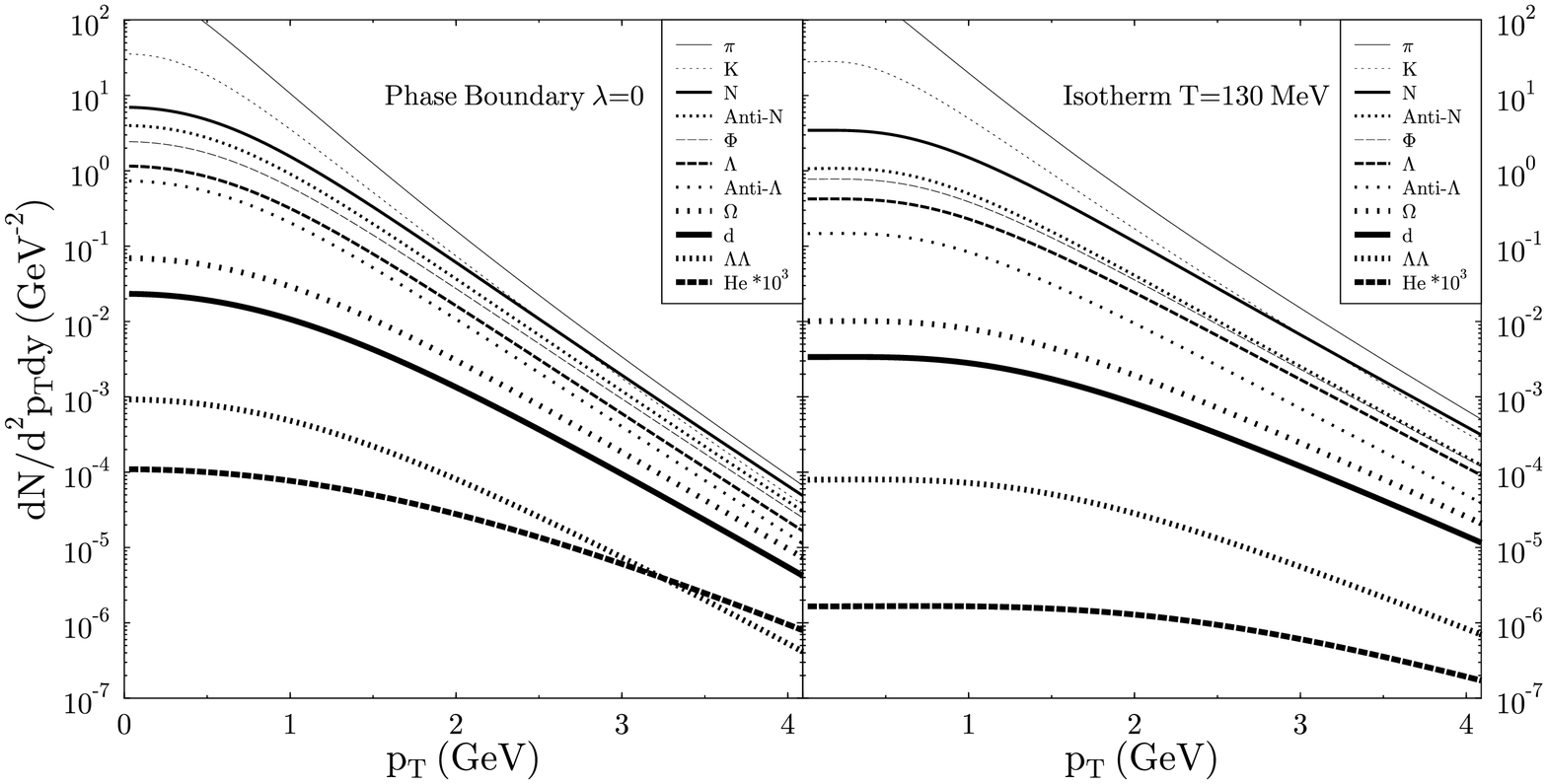,height=8cm,width=13cm}}}
\vspace*{1cm}
\caption{Transverse momentum spectra of direct thermal hadrons at midrapidity
(central $Au+Au$ collisions at RHIC), as calculated on the $\lambda=0$ and
$T=130$~MeV hypersurfaces, respectively. A ``wounded nucleon'' distribution
has been employed for the initial transverse energy and baryon density.}
\label{dndptRHICwn}
\vspace{.5cm}
\end{figure}
However, if the initial transverse energy and baryon density distributions
exhibit smaller gradients, this kink in the spectra is washed out, cf.\
Fig.~\ref{dndptRHICwn}. For the ``wounded nucleon'' distribution,
eq.~(\ref{wn_distrib}),
the very fast tails in the velocity distribution on the hypersurface are
absent, thus reducing the number of hadrons at $p_T>2$~GeV.
Figures~\ref{dndptRHIC} and \ref{dndptRHICwn} show the possible range 
of collective particle production at high transverse momenta.
On the other hand, we have found that the two initial density distributions
lead to {\em average} transverse momenta that differ by $\le10\%$.
Therefore, we present results for $\langle p_T\rangle$ only for the sharp
density profile.

\subsection{Mean Transverse Velocities and Momenta}
\begin{figure}[hbp]
\centerline{\hbox{\epsfig{figure=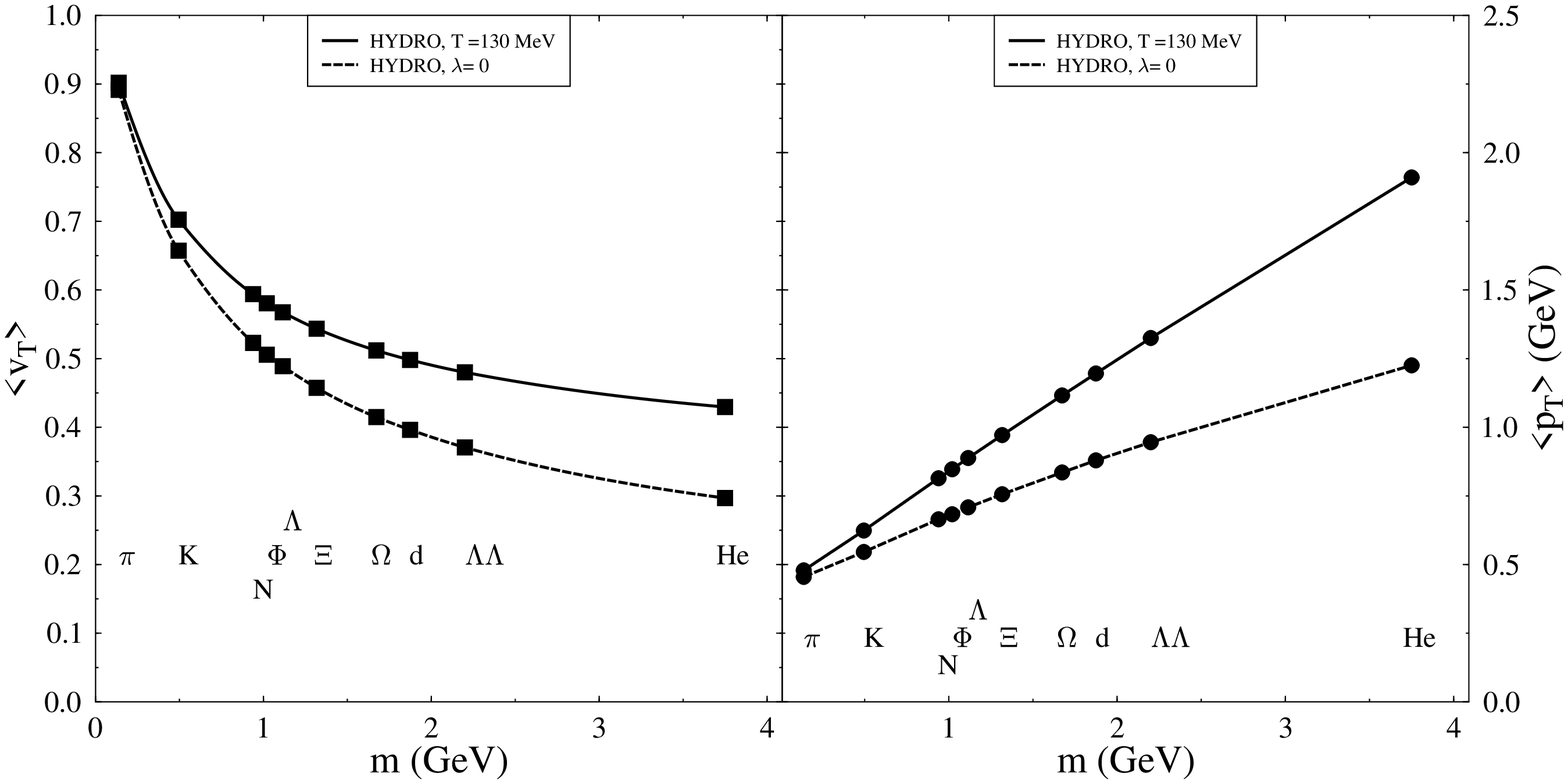,height=8cm,width=13cm}}}
\vspace*{1cm}
\caption{Average transverse momenta and velocities (at midrapidity) of direct
thermal hadrons on the boundary between mixed phase and pure hadron phase,
$\lambda=0$, and on the $T=130$~MeV isotherm
(for central $Pb+Pb$ collisions at SPS).}
\label{meanptSPS}
\vspace{.5cm}
\end{figure}
\begin{figure}[htp]
\centerline{\hbox{\epsfig{figure=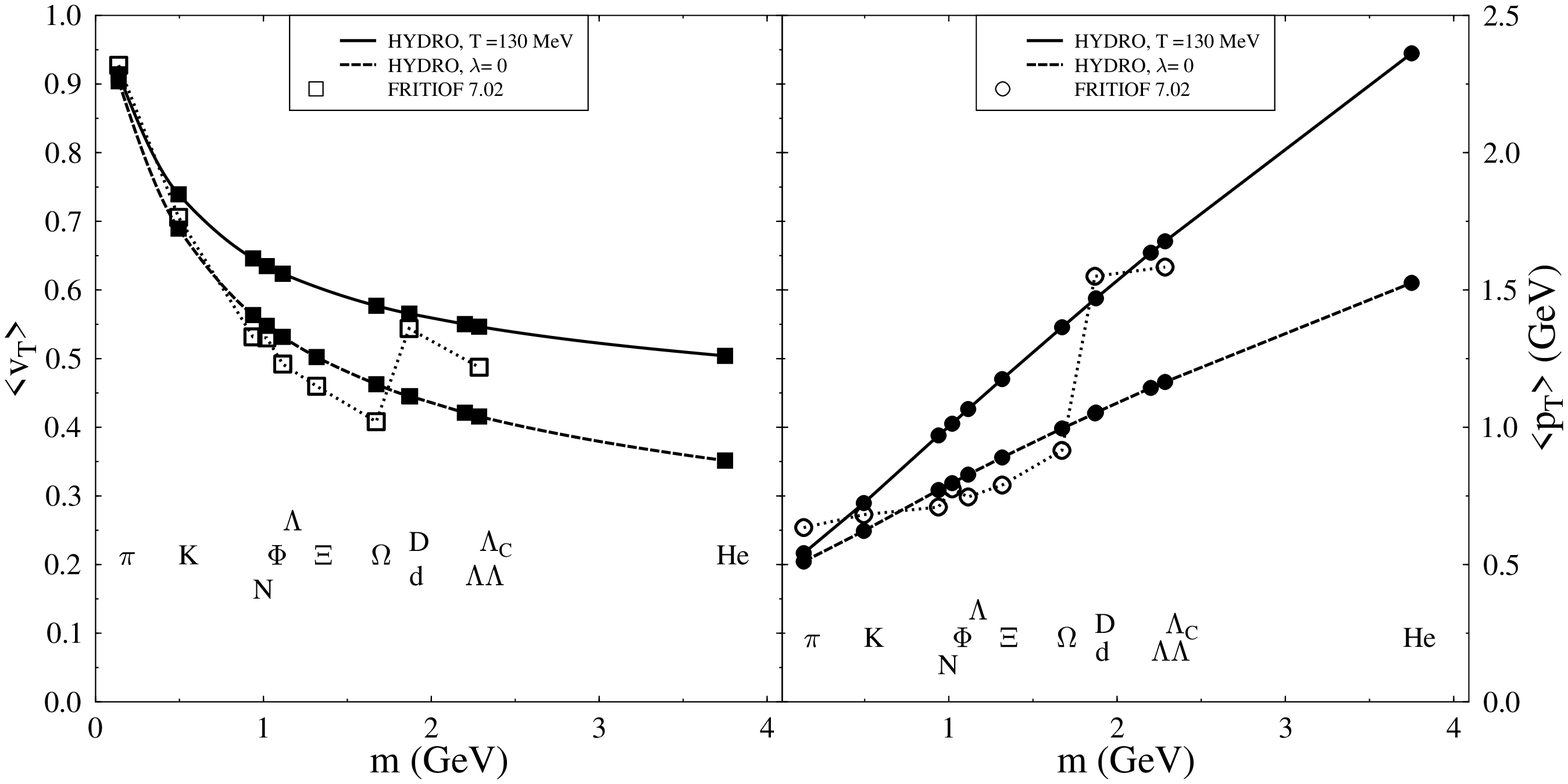,height=8cm,width=13cm}}}
\vspace*{1cm}
\caption{Same as fig.\ \ref{meanptSPS} but for
central $Au+Au$ collisions at RHIC. Open symbols depict the predictions
of {\small FRITIOF~7.02}.}
\label{meanptRHIC}
\vspace{.5cm}
\end{figure}

  
Figures \ref{meanptSPS} and \ref{meanptRHIC} depict the average transverse
momenta and velocities
of various hadrons at midrapidity, as calculated from the $p_T$ distributions:
\be
(\langle p_T\rangle_i,\langle v_T\rangle_i) = \frac{\int d^2p_T\,
(p_T,p_T/p^0) \frac{dN_{i}}{d^2p_T dy} }
{\int d^2p_T\,\frac{dN_{i}}{d^2p_T dy}}\quad.
\ee
One observes that due to the collective transverse expansion, $\langle
p_T\rangle=m\langle\gamma_T v_T\rangle$ increases linearly with the particle
mass\footnote{Note, however, that $\langle\gamma_T v_T\rangle$ does in
principle depend on the particle mass, since it includes flow as well as
thermal velocity components, cf.\ left panel of Fig.~\ref{meanptSPS}.
Thus, the result $\langle p_T\rangle\propto m$ is not trivial.}.
The $\langle p_T\rangle_\pi=470$~MeV of pions at SPS is slightly
overpredicted, NA49 measured $418\pm30$~MeV~\cite{NA49h-}. However, this
is not astonishing since we have employed the Boltzmann approximation,
and have also not included pions from
post-freeze-out resonance decays. For all other (heavier)
hadrons these effects on $\langle p_T\rangle$ are less significant.

On the other hand, $\langle v_T\rangle$
saturates with increasing $m$, since the thermal velocities $\sim
\sqrt{T_{fo}/m}$ become negligible and $\langle v\rangle\approx v_{flow}$. 
Thus, clusters
($d$, $He$, ...) provide the opportunity to determine the transverse
flow velocity {\sl directly} \cite{RaffiPRL,StoOG}, while the thermal
averaging cannot be neglected for nucleons and especially for pions
and kaons. To extract the collective flow from their
$\langle p_T\rangle$ or $\langle v_T\rangle$ requires
the knowledge of
the freeze-out temperature, and moreover might be distorted by
decays of resonances~\cite{vTExp,SPSvTth}. The average transverse flow
velocity in $Pb+Pb$ at SPS is $\approx0.3$ on the boundary between
mixed and hadronic phase (i.e.\ $\lambda=0$), and
increases to $\approx0.4$ on the $T=130$~MeV isotherm. This is consistent
with other fluid dynamical calculations~\cite{JSol,SHR,BSch,HungS} and
numbers extracted from fits of transverse momentum
distributions~\cite{NA49phi,BMS,vTExp,SPSvTth}. Our analysis does not support
freeze-out temperatures of less than $100$~MeV at SPS~\cite{Nix},
at least not at midrapidity.

Although the entropy per net baryon at RHIC is five times larger than at SPS,
the transverse flow velocity increases only slightly. (In particular, the
$\langle p_T\rangle$ of pions almost stagnates from SPS to RHIC.)
This is due to the fact that the transverse flow cannot increase strongly
in the expansion of the mixed phase, since the velocity of sound is very
small~\cite{RiGy,RiGy2,SoftTr}. Thus, hydrodynamics
predicts that the transverse flow velocity at RHIC is comparable to that
at SPS, if the EoS exhibits a first-order phase transition.
If freeze-out at RHIC
occurs as early as on the $\lambda=0$ hypersurface, the transverse
flow velocity could even be smaller than at SPS, cf.\ Figs.~\ref{meanptSPS}
and~\ref{meanptRHIC}.

It is also interesting to note that both at SPS and RHIC a significant
part of the transverse flow is already produced within the QGP and (to
a smaller part also within) the mixed phase, compare the dashed
curves in figs.~\ref{meanptSPS} and \ref{meanptRHIC} with the solid lines.
This behavior is different to that observed within the
{\small RQMD~2.3} model, where the transverse flow of baryons emerges mainly
from the late stage of the reaction~\cite{Hecke}. The reason is that within
{\small RQMD~2.3} the hadron gas is not in kinetic equilibrium at times
$t\stackrel{~}{<}5$~fm, while we assume early kinetic equilibrium of the
QGP in the central region, $\tau_i=1$~fm.

The flow of the $\phi$ meson is closely related to this point.
Even if $\phi$ mesons do not rescatter with surrounding hadronic or mixed
phase matter, their $\langle p_T\rangle$, $\langle v_T\rangle$ or inverse
slope in AA collisions should be affected by collective flow
if a QGP with $s$ quarks in local thermal equilibrium is created. This is
due to the fact that the transverse momenta of $s$ quarks are subject
to a Doppler shift according to the collective flow velocity built up
in the QGP phase. According to Fig.~\ref{meanptSPS}, $\langle p_T
\rangle_\phi\approx680$~MeV and $\langle v_T\rangle_\phi\approx0.5$ at SPS,
even if the $\phi$'s decouple from collective flow immediately at the boundary
between mixed and hadronic phase. Indeed, preliminary NA49 data \cite{NA49phi}
show that the $\phi$ roughly fits into the $p_T(m)$ systematics predicted by
hydrodynamics, as already indicated by Fig.~\ref{dndptSPS}.
Note that the ``conventional'' mechanism to generate $p_T$ in cascade
models, i.e.\ resonance decay or color flux-tube fragmentation followed by
additional rescattering with the measured free (or additive quark model)
cross section, apparently works for $\pi$, $K$, $N$, $\Lambda$ but not
for the $\phi$~\cite{URQMDphi} (cf.\ also the discussion in ref.~\cite{HSS}).

For comparison, we have also calculated the average transverse momentum
and velocity of direct hadrons\footnote{Clusters ($d$, $\Lambda\Lambda$,
and $He$) are not computed within this model.} (i.e.\ resonance decays were
also switched off) at RHIC within {\small FRITIOF~7.02}~\cite{FRI}, cf.\
fig.~\ref{meanptRHIC}. In this string model, secondary hadrons do not
rescatter.
Thus, collective transverse flow does not develop. One observes that even
though hard scatterings are taken into account, the $\langle p_T\rangle$ of
$N$, $\phi$, $\Lambda$, and $\Omega$ is not higher than in case of
hydrodynamic
expansion with freeze-out on the $\lambda=0$ hypersurface. This difference
would be even bigger if collective flow set in earlier (remember that
we assume $\tau_i=0.6$~fm).

Only the $\langle p_T\rangle$ of
charmed hadrons, $D$ meson and $\Lambda_C(2285)$ baryon\footnote{Pions and
kaons are not very sensitive to the dynamical evolution and
are left out of the discussion for the moment.}, is strongly enhanced by
(semi-)hard scatterings, and is close to the solid line in
fig.~\ref{meanptRHIC}. These very heavy hadrons are produced by fragmentation
of semi-hard jets and acquire high transverse momentum.
Thus, in {\small FRITIOF~7.02} the $\langle p_T\rangle$
of various hadrons does not increase linearly with mass but depends
strongly on the sensitivity to (semi-)hard scatterings, and thus to the
flavor content. In contrast, if $D$ mesons thermalize~\cite{Svetitsky} and
follow the collective transverse flow, their $\langle p_T\rangle$ is
``pulled'' back onto the linear $\langle p_T\rangle(m)$ systematics.

\section{summary}
In this paper we have investigated the collective evolution of the
central rapidity region with initial conditions as expected for heavy-ion
collisons at CERN-SPS and BNL-RHIC energies. Even in the case where the 
phase transition (assumed to be of first order) from the
plasma of $u$, $d$, $s$ quarks and gluons to the hadron gas proceeds
in local thermodynamical equilibrium, collective transverse expansion and
equilibration of heavy mesons and (anti-)baryons reduces the hadronization
time to $\le3R_T$ (for $Au$ nuclei). Nevertheless, at RHIC, the lifetimes of
the QGP and the mixed phase are significantly longer than at SPS.
However, at SPS the midrapidity region evolves further into the hadronic
phase, which occupies a larger space-time volume than the QGP or mixed phase.
As a consequence, the transverse flow velocity and the mean transverse
momentum of the particles increase substantially during this stage of the
reaction. Without this ``hadronic afterburner'' the hydrodynamical model as
discussed in this paper would not describe the measured
single-particle transverse momentum distributions.

At RHIC, one might expect a higher freeze-out temperature than at SPS,
since the baryon density is smaller. If, for example, freeze-out occurs
directly after hadronization is complete (i.e.\ on the $\lambda=0$
hypersurface), the average transverse flow velocity turns out to be similar
or even smaller than at SPS.

We have calculated the transverse momentum distributions of
various hadrons up to $p_T=4$~GeV on the $T=130$~MeV isotherm and on the
$\lambda=0$ hypersurface for SPS and RHIC, respectively. Large gradients
in the initial transverse density distributions (e.g.\ a sharp step function)
can strongly enhance particle production at high $p_T$. The recently
observed $\pi^0$ distribution (up to $p_T=4$~GeV)
in $Pb+Pb$ at $\sqrt{s}=18A$~GeV can in
principle be described by a hydrodynamic expansion. However,
the high-$p_T$ tails of the hadron distributions (above $p_T=2$~GeV) are
sensitive to the initial pressure gradients.

Within hydrodynamics the mean transverse momentum
of hadrons increases
linearly with the particle mass and is independent of other quantum
numbers (provided the corresponding hadron species 
is in local thermal equilibrium).
Heavy hadrons and clusters, if identified, allow for a direct measurement of
the collective transverse expansion velocity, without requesting knowledge
of freeze-out temperatures.

Finally, we have compared the fluid-dynamical results to those
obtained within the string
model {\small FRITIOF~7.02}. Due to the increasing ``sensibility'' on
(semi-)hard processes, the $\langle p_T\rangle$ of hadrons with
$c$ quarks increases more strongly than the
$\langle p_T\rangle$ of hadrons with only light (anti-)quarks. Consequently,
there is a strongly nonlinear relation between the mean transverse momentum
and the particle mass within this model.
\acknowledgements
We thank L.\ Gerland, M.\ Gyulassy, G.\ Kunde, T. Peitzmann, K.\ Redlich,
and C.\ Spieles for many helpful discussions,
C.\ Bormann, J.\ G\"unther, G.\ Roland, and C.\ Struck for providing the
NA49 data, and T. Peitzmann for communicating the preliminary WA98 data.
A.D.\ gratefully acknowledges a postdoctoral fellowship by the German
Academic Exchange Service (DAAD).

\end{document}